\documentclass[a4paper,11pt]{article}
\usepackage[utf8]{inputenc}
\usepackage{amsmath, amssymb, amsthm}
\usepackage{graphicx}
\usepackage{geometry}
\usepackage{hyperref}
\usepackage{enumitem}
\usepackage{url}

\usepackage{booktabs}
\usepackage{subcaption}

\title{Skewness-Kurtosis: small samples and power-law behavior
}

\author{
  Carlo De Michele\thanks{Politecnico di Milano, Dipartimento di Ingegneria Civile e Ambientale (carlo.demichele@polimi.it)}
  \and
  Samuele De Bartolo\thanks{Department of Engineering for Innovation, EUropean Maritime Environmental Research (EUMER), Università del Salento, via per Monteroni S.P.6, Lecce, 73100, Italy (samuele.debartolo@unisalento.it)}
}

\date{}

\begin{document}

\maketitle

\begin{abstract}

Skewness and kurtosis are fundamental statistical moments commonly used to quantify asymmetry and tail behavior in probability distributions. Despite their widespread application in statistical mechanics, condensed matter physics, and complex systems, important aspects of their empirical behavior remain unclear — particularly in small samples and in relation to their hypothesized power-law scaling. In this work, we address both issues using a combination of empirical and synthetic data. First, we establish a lower bound for sample kurtosis as a function of sample size and skewness. Second, we examine the conditions under which the 4/3 power-law relationship between kurtosis and skewness emerges, effectively extending Taylor's power-law to higher-order moments. Our results show that this scaling behavior predominantly occurs in data sampled from heavy-tailed distributions and medium/large sample sizes.

\end{abstract}

\section{Introduction}
\label{intro}
Skewness and kurtosis are dimensionless, normalized statistical moments of the third and fourth order, respectively. Skewness quantifies the asymmetry of a probability distribution around its mean, indicating whether the data are skewed toward higher or lower values. Kurtosis, on the other hand, measures the "tailedness" of the distribution, or the propensity of the data to exhibit extreme deviations from the mean. Together, these moments provide information about the shape characteristics of distributions beyond mean and variance.
These metrics are widely used in a broad range of scientific disciplines, including turbulence \cite{Busseetal23}, oceanography \cite{Sura2008,HUGHES2010,Kong25}, climatology \cite{Tamarin-Brodsky22}, meteorology \cite{Maurizi1999}, hydrology \cite{Cugerone15,Cugerone17}, plasma physics \cite{Sattin09,Garcia12}, seismology \cite{Cristelli12}, and finance \cite{Cristelli12,KARAGIORGIS22,Karagiorgisetal24}. 

Understanding the relationship between skewness and kurtosis has important practical implications. In fields such as finance, geophysics, environmental science, and engineering, data distributions often exhibit asymmetry (skewness) and heavy tails (kurtosis), which are indicators of rare but impactful events. A quantitative relation between these two moments can help in characterizing the nature of extreme events, assessing risk, and improving anomaly detection. For example, in financial time series, deviations from normality are critical for volatility modeling and option pricing \cite{Bonato22, FOGLIA2025}. In climate, ecological, and human systems, skewness–kurtosis patterns can provide early warnings of regime shifts or critical transitions \cite{Dakos2024}. 
In addition, interest in evaluating skewness and kurtosis in small samples is motivated by the need for localized assessments in space or time. 
Therefore, a structural understanding of how these moments scale with one another can improve modeling accuracy and decision-making in real-world systems.
However, the interpretation of skewness and kurtosis must be approached with caution in certain contexts. In particular, for heavy-tailed distributions where $P(X>x) \propto x^{-\alpha}$, the statistical moments of order $\ge \alpha$ diverge. Specifically, the third moment (skewness) does not exist for
$0 < \alpha < 3$, and the fourth moment (kurtosis) does not exist for $0 < \alpha < 4$. In such cases, sample-based estimates can be misleading. Nonetheless, plotting sample values in the skewness–kurtosis plane remains a common empirical strategy to infer the underlying probability distribution—often represented as a geometric locus (point, curve, or region) that best fits the data \cite{Cugerone17}.
The sample skewness and kurtosis are functions of the sample size and bounded. In particular, Johnson and Lowe \cite{Johnson79}, Wilkins \cite{Wilkins44} and Kirky \cite{Kirby74} provided the upper limit of the sample skewness as a function of the sample size. Similarly, Johnson and Lowe \cite{Johnson79}, Dalén \cite{Dalén87} gave the upper limit of the sample kurtosis as a function of the sample size. More recently, Sharma and Bhandari \cite{Sharma15} provided the upper limit of the sample kurtosis as a function of the sample skewness and sample size, while reported the lower limit as a function of the sample skewness. However, the dependence of the upper limit on skewness and sample size might suggest that the lower limit could also be influenced by the sample size. 
Concerning the relation between sample kurtosis and skewness, Cristelli et al. \cite{Cristelli12} proposed the 4/3 power-law as universal relationship. Using synthetic data, Celikoglu and Tirnakli \cite{Celikoglu15} criticized this result, arguing that the 4/3 power-law is not a universal relation, and originates only due to the small sample size of data. In literature, a power-law relationship between variance and mean has been proposed by Taylor \cite{TAYLOR1961}, and referred as Taylor Power-Law (TPL). It has been used to explain the population sizes or population densities of species distributed in space and time in ecology,
and has been observed in a wide range of other taxa, communities, and even non-ecological systems \cite{Taylor19}, such as astronomy, finance, and geophysics, including phenomena like earthquakes, precipitation, and braided rivers \cite{DEBARTOLO2024,DeBartoloetal2022}.
Cohen and Xu \cite{CohenXu15} investigated the Taylor power-law using synthetic samples, demonstrating that samples drawn from skewed probability distributions with four finite moments exhibit a power-law relationship between variance and mean. In this respect, a power-law relationship between skewness and kurtosis can be considered as an extension of the Taylor Power-Law (a second-order theory) to a fourth-order theory.
In Section 2, firstly, we report the existing knowledge about 1) the behavoir of sample skewness and kurtosis as function of the sample size, and 2) the emergence of power-law between skewness and kurtosis, and 3) we provide a criterion to detect the emergence of the power-law. Then, we address the two key issues, i.e., the behavoir of sample skewness and kurtosis for small samples and the emergence of the power-law, using both observational and synthetic datasets (extracted from discrete and continuous distributions). In Section 3, we describe the observed data used. In Section 4, firstly, we analyze the behavior of skewness and kurtosis in small samples ($n = 4$–$9$), establishing a lower bound for sample kurtosis as a function of skewness and sample size; secondly, we investigate the emergence of a power-law relationship between skewness and kurtosis. Finally, in Section 5, we provide our conclusions.

\section{Methodology}
\label{sec:1}
\subsection{Factual Knowledge}
Sample skewness and kurtosis (indicated respectively as $S$ and $K$) can be written as
\begin{eqnarray}\label{equation1}
S=\frac{m_3}{m_2^{3/2}}\lessgtr 0  \qquad     
K=\frac{m_4}{m_2^{2}}>0
\end{eqnarray}
where $m_r=\frac{1}{n} \sum_{i=1}^{n}(x_i-\bar{x})^r$
is the $r$th central moment and $\bar{x}=\frac{1}{n}\sum_{i=1}^n x_i$ is the arithmetic
mean of a sample $(x_1, x_2, ..., x_n)$ of size $n$. 
If the sample is symmetric with respect to the mean $\bar{x}$, then $m_3=0$ and also the sample skewness $S=0$.
$S$ and $K$ have bounds depending on the sample size $n$. 

Johnson and Lowe \cite{Johnson79}, elaborating the term $m_2^2$, showed that 
\begin{equation}\label{equation21}
K \le n
\end{equation} and then 
\begin{equation}\label{equation20}
|S|\le \sqrt{n-1}
\end{equation}

Wilkins \cite{Wilkins44} and Kirby \cite{Kirby74} proved that

\begin{equation}\label{equation2}
|S|\le \frac{n-2}{\sqrt{n-1}}
\end{equation}
while Dalén \cite{Dalén87}
\begin{equation}\label{equation3}
K\le \frac{n^2-3n+3}{n-1}
\end{equation}
Sharma and Bhandari \cite{Sharma15} using the Newton's disequality showed that 
\begin{equation}\label{equation4}
1+S^2\le K \le \frac{1}{2}\frac{n-3}{n-2}S^2+\frac{n}{2}
\end{equation}
Eq.(\ref{equation4}) provides lower and upper limits of $K$ given $S$ and $n$, recognizing that skewness and kurtosis are not independent each other. Note that in Eq.(\ref{equation4}), the upper bound depends on $S$ and $n$, while the lower bound only depends on $S$. The left-hand side disequality, that is, $K\ge S^2+1$, is also known as the Pearson lower bound \cite{Pearson16}. Approximating in sample skewness and kurtosis (Eq.\ref{equation1}), the sums with the term corresponding to the maximum value, Cristelli et al.  \cite{Cristelli12} proposed the following $4/3$ power-law as universal relationship between $K$ and $S$
\begin{equation}\label{equation5}
K \simeq n^{1/3} S^{4/3}
\end{equation}
On the contrary, using synthetic data, Celikoglu and Tirnakli \cite{Celikoglu15} criticized Cristelli et al. \cite{Cristelli12} arguing that Eq.(\ref{equation5}) is not a universal relation, and originates only due to the small number $n$ of data in the datasets considered.
In the literature, a power-law relationship has been observed between the first two statistical moments, variance $\sigma^2$ and mean $\mu$, $\sigma^2=b_1\mu^{b_2}$, where $b_1$ is a constant, and $b_2$ is a scaling exponent or ‘index of aggregation’. This relationship, proposed by Taylor \cite{TAYLOR1961}, is known as the Taylor Power-Law (TPL).
A power-law relationship between skewness and kurtosis like Eq.(\ref{equation5}) can be considered an extension of the Taylor Power-Law (a second-order theory) to a fourth-order theory.

\subsection{Present Development}
From Eq.(\ref{equation1}), by writing $m_{2}=\sqrt{\frac{m_4}{K}}$ and substituting this in the expression of $S$, we write $S = \frac{m_3}{m_4^{3/4}}K^{3/4}$. After further manipulation, this leads to the relation 
\begin{equation}\label{equation6}
K \cdot R= n^{1/3}S^{4/3} 
\end{equation}
where $R=\frac{(\sum_{i=1}^{n}(x_i-\bar{x})^3)^{4/3}}{\sum_{i=1}^{n}(x_i-\bar{x})^{4}}\geq 0$ is a scale-invariant ratio, depending on the sample. If the sample is symmetric with respect to $\bar{x}$, then the numerator of $R$, $\sum_{i=1}^{n}(x_i-\bar{x})^3)$ is zero, as well as $m_3=0$, and $S=0$. 

In the case $R>0$, Eq.(\ref{equation6}) can be rearranged as 
\begin{equation}\label{equation7}
K =\frac{1}{R} n^{1/3}S^{4/3} 
\end{equation}

The emergence of the 4/3 power-law relationship between sample kurtosis \( K \) and sample skewness \( S \) is closely tied to the behavior of the scale-invariant ratio \( R \). Specifically, when \( R \) exhibits low variability across samples—that is, it remains approximately constant—the relation \( K \propto S^{4/3} \) becomes apparent. In such cases, the consistency of \( R \) supports the emergence of the power-law behavior. Vice versa, if \( R \) varies significantly across samples, then the \( 4/3 \) power-law relationship between \( K \) and \( S \) may break down, or become less apparent. In such cases, the variability of \( R \) introduces deviations from the 4/3 scaling behavior, leading to a dispersion of data points around the theoretical power-law curve.

If the samples contain a few extreme values that dominate both the numerator and denominator of \( R \), then \( R \approx 1 \). In this case, Eq.~(\ref{equation7}) with \( R = 1 \) recovers the empirical relationship observed by Cristelli et al.~\cite{Cristelli12}.
Equation~(\ref{equation7}) thus represents a structural relation between \( K \) and \( S \), with the \( 4/3 \) power law emerging under specific statistical conditions.

To assess the emergence of power-law behavior between $K$ and $S$, we propose the use of $P_{R|S>s_q}$ the conditioned probability distribution of $R$, given $S$ greater than a threshold $s_q$, $S>s_q$, where $s_q$ is the $q$-quantile of $S$. 
This criterion can be written as 
\begin{equation}\label{equation8} 
P_{R|S>s_q}(1-\epsilon\le R\le 1+\epsilon) \approx 1
\end{equation}
in which $\epsilon \in \mathbb{R}^{+}$ is a small tolerance parameter. Here, as rule of thumb, we have used $\epsilon = 0.05$. 
If this condition is satisfied for
small values of $q$, then the $4/3$ power-law between $K$ and $S$ is well representative for all the samples, viceversa if this condition happens for a $q->1$ then it is descriptive only of some samples.

\section{Data}
We have considered three datasets: i) air humidity data (in \%) collected in Toulouse (France), each 15 min, in the period 21/12/2022--04/02/2025, available at \url{https://data.toulouse-metropole.fr/explore/dataset/}; 
ii) Rakaia and Brahmaputra data consisting of the number of wet channels, for each cross section measured for the entire longitudinal development of the rivers \cite{DeBartoloetal2022,Rizzelloetal24};
iii) precipitation data (in 0.1 mm) collected at Linkoeping-Malmslaett (Sweden), with daily sampling, in the period 01/12/1858--31/12/2024, available at \url{https://www.ecad.eu/dailydata/index.php}. For this dataset, we considered daily precipitation $>0$. 

\section{Results}
We evaluated the sample skewness and kurtosis of datasets for different sample sizes $n$, creating partitions 
$\Delta=N/n$ of the total database size 
$N$. The rationale of this way of doing (used in time and/or space domain) is to investigate hidden structure, anomalies, and scale-dependent or time-varying properties of a dataset/ phenomenon. Warsza and Korczynski \cite{Warszaetal15} also reported reasons for dealing with small measurement samples.  

\begin{figure*}[htbp]
    \centering
    \begin{subfigure}[b]{0.405\textwidth}
        \includegraphics[width=\textwidth]{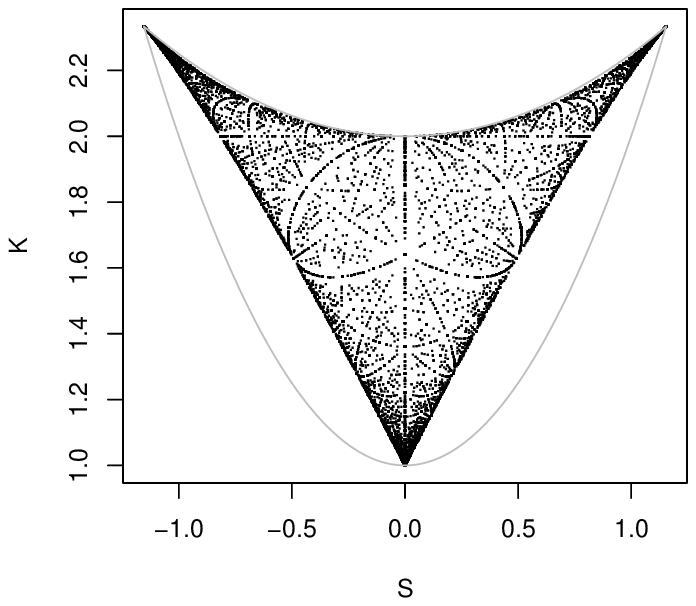}
        \caption{$n=4$, Toulouse humidity data.}
    \end{subfigure}
    \hfill
    \begin{subfigure}[b]{0.405\textwidth}
        \includegraphics[width=\textwidth]{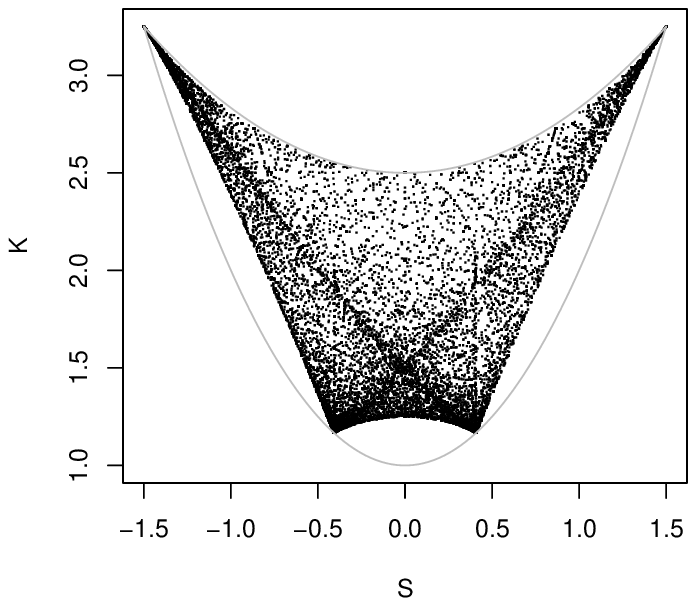}
        \caption{$n=5$, Toulouse humidity data.}
    \end{subfigure}
    
    \begin{subfigure}[b]{0.405\textwidth}
        \includegraphics[width=\textwidth]{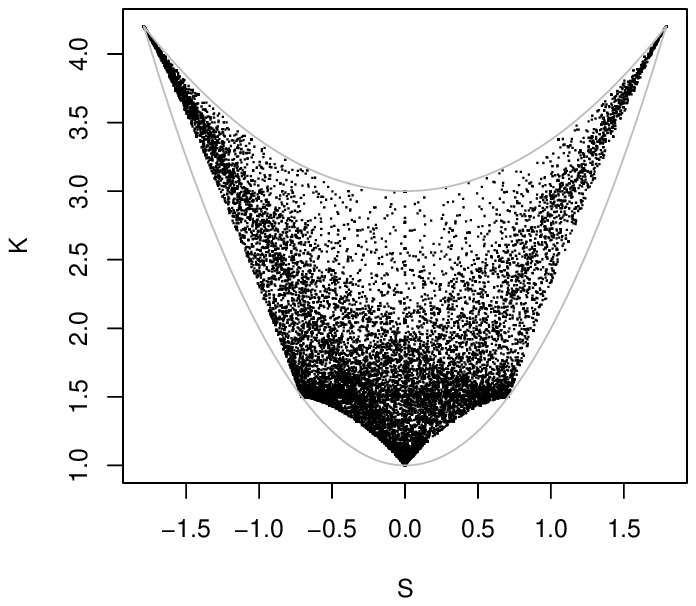}
        \caption{$n=6$, Toulouse humidity data.}
    \end{subfigure}
    \hfill
    \begin{subfigure}[b]{0.405\textwidth}
        \includegraphics[width=\textwidth]{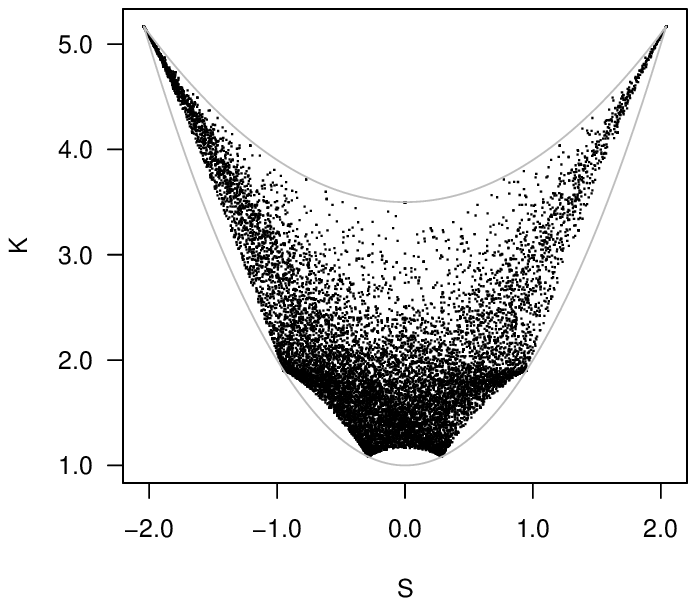}
        \caption{$n=7$, Toulouse humidity data.}
    \end{subfigure}

    \begin{subfigure}[b]{0.405\textwidth}
        \includegraphics[width=\textwidth]{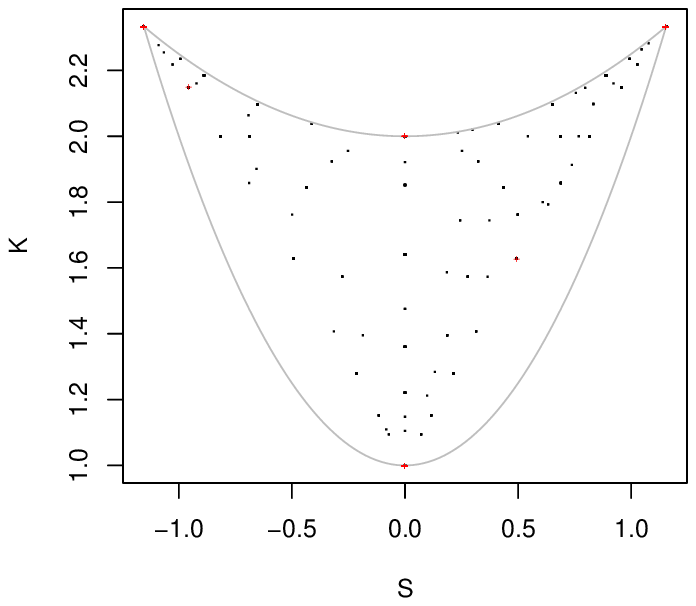}
        \caption{$n=4$, Rakaia (red) \& Brahmaputra (black) river data consisting of the number of wet channels, for each cross section.
        }
    \end{subfigure}
    \hfill
    \begin{subfigure}[b]{0.405\textwidth}
        \includegraphics[width=\textwidth]{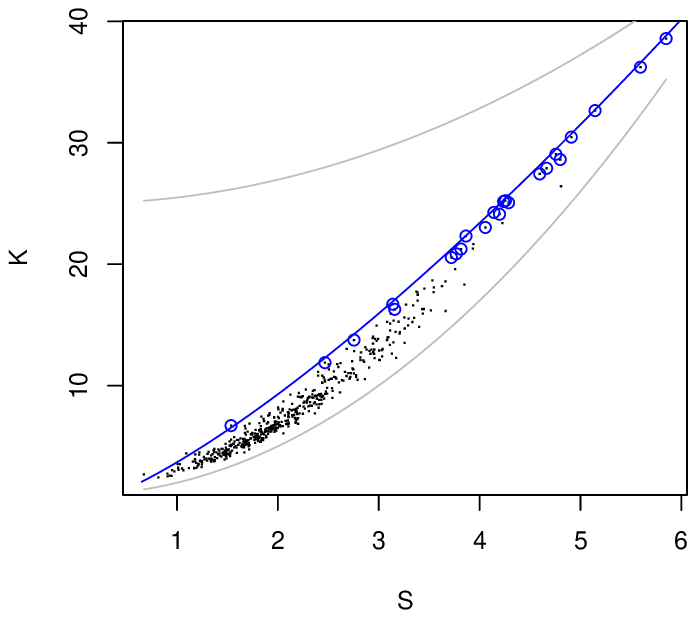}
        \caption{$n=50$, Linkoeping-Malmllaett daily precipitation data. Blue line is Eq.(\ref{equation7}), blue circles are the couples with a $R$ factor in [0.95, 1.05].} 
    \end{subfigure}

    \caption{\label{fig:1}Values of skewness and kurtosis for different sample sizes $n$ and typologies of data. The grey lines represent the upper and lower limits given by Eq.(\ref{equation4}).}
\end{figure*}

In Fig.\ref{fig:1}, the sample skewness and kurtosis are shown using the following sizes: $n=4$ panels a) and e), $n=5$ panel b), $n=6$ in panel c), $n=7$ in panel d), and $n=50$ in panel f). In addition, in Fig.\ref{fig:1}, the upper and lower limits given by Eq.(\ref{equation4}) (grey lines) are also reported.
From Fig.(\ref{fig:1}a-e), it is possible to note that i) for small values of $n$, while the upper bound well limits the sample values, the data occupies geometrical domains, which depend on $n$, implying that also the lower limit in Eq.(\ref{equation4}) should be a function of $n$; ii) in Fig.(\ref{fig:1}a), it seems that inside the domain, a deltoid shape, data describes geometrical patterns; iii) in Fig.(\ref{fig:1}f) the high values of skewness and kurtosis appear to be aligned along a power-law function.  

In the next, in order to investigate the skewness-kurtosis behavior in small samples and the emergence of a power-law relationship, we have generated synthetic samples, $N=10^8$ in length, partitioned in different sizes $n$ (small values $n=4-9$, medium values $n=100$, large values $n=1000$) extracted from different probability distributions, both continuous, including Gaussian(mean=0, sd=1), Exponential(rate=1), Lognormal(meanlog=0, sdlog=1),\\
Gamma(shape=2, rate=1), Pareto(shape=5, scale=1), and discrete, including Binomial(trials= 20, p=0.8), Negative Binomial(trials=20, p=0.8), Poisson(rate=20), Geometric(p=0.8), Zipf(shape=5, xmin=1). 

\begin{figure*}[b]
{\includegraphics[width=0.8\textwidth]{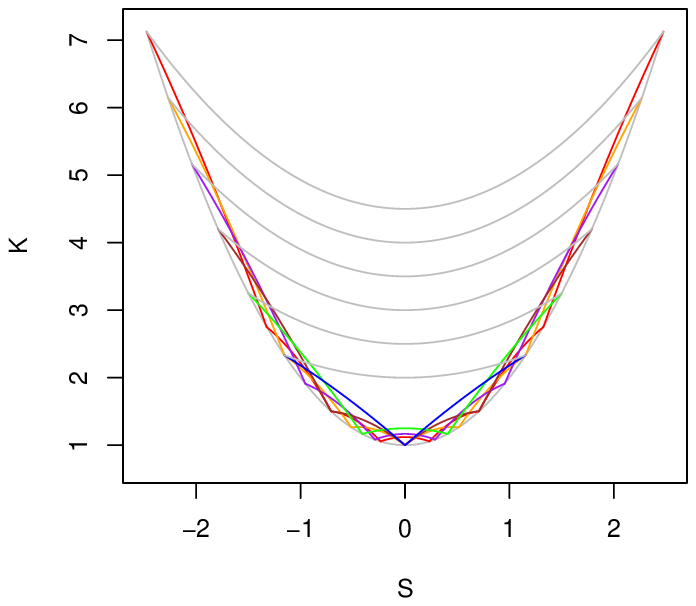}}
\caption{\label{fig:2} Lower limit of sample kurtosis as function of $n$: blue $n=4$, green $n=5$, brown $n=6$, purple $n=7$, orange $n=8$, red $n=9$. In grey are reported also the limits given by Eq.(\ref{equation4}) for $n=4,..,9$.}
\end{figure*}

Firstly, using small samples, i.e., $n=4-9$, we have empirically determined the lower bound of kurtosis as a function of skewness and $n$. Since the number of samples is $N/n\sim 10^7$, the probability of finding a kurtosis value smaller than the lower bound is $< 10^{-7}$. This makes such an occurrence extremely unlikely, thereby indicating that the estimate of the lower bound is highly precise. We find that the lower bound is well represented by a parabolic law $K=aS^2+bS+c$, where the coefficients $a,b,c$ are functions of $n$ and depending on the $S$ ranges.
In Table 1, for $n$ that varies between $(4,9)$, the coefficients $(a,b,c)$ and the range of $S$ are given for the left side of the lower limit. The right side is not reported because the lower bound is symmetric with respect to the y-axis.
In Fig.\ref{fig:2}, we report the lower limit of the sample kurtosis as a function of $S$ and $n$: blue $n= 4$, green $n = 5$, brown $n = 6$, purple $n = 7$, orange $n = 8$ and red $n = 9$. In Fig.\ref{fig:2}, we also provide the limits given in Eq.(\ref{equation4}). We find that 1) the lower and upper limits, for fixed $n$, are invariant with the distribution considered, while the occupancy of the skewness-kurtosis couples,  within the domain, is a function of the distribution selected. Although continuous distributions tend to occupy all the space within the domain, discrete distributions exhibit patterns inside the domains. This is particularly evident in case $n=4$, given in Fig.\ref{fig:3}, obtained using Poisson(rate=20). In this domain, with a deltoid shape, it is possible to see clear geometrical patterns characterized by self-similar structures. Remarkable symmetry with respect to the $S = 0$ axis is observable.
The deltoid is a fractal set whose dimension, estimated using a standard direct box counting procedure \cite{debartolo06}, is $1.798\pm 0.006$ with $R^2=0.9997$. 2) Increasing $n\gg10$ the lower limit tends to be well represented by the Pearson limit, i.e., $K=S^2+1$. 
  
In Fig.\ref{fig:4}, we report the empirical conditioned cumulative distribution of $R$, obtained using $\Delta=10^8/n$ samples, and different $q=0.9, 0.95, 0.99, 0.999$, 0.9995, $0.9999, 0.99995$, under four cases: a) Gaussian(mean=0, sd=1) samples of size $n=100$; b) Exponential(rate=1) with $n=1000$; c) Lognormal(meanlog=0, sdlog=1) with $n=100$; and d) Pareto(shape=5, scale=1) samples with $n=1000$. From Fig.\ref{fig:4}, in cases a) and b) the condition given by Eq.(\ref{equation8}) is not satisfied and the power-law behavior does not emerge, while in cases c) and d) the condition given by Eq.(\ref{equation8}) is satisfied and the $4/3$ power-law behavior is found. In Table \ref{tab:table2}, we report for the distributions considered and the sample size $n$, in which cases the power-law emerges. While for $n=4$ the power-law does not emerge, as expected for the bounded behavior of $S$ and $K$, for $n=100$ and $n=1000$, we found that for $n=100$, it emerges
for continuous distributions in the case of Exponential, Lognormal, Gamma and Pareto distributions, and for discrete distributions in the case of Geometric and Zipf. For $n=1000$, it emerges for continuous distributions in the case of Lognormal and Pareto distributions and for discrete distributions in the case of Zipf. As a general comment, it is possible to say that the power-law behavior emerges in the case of samples extracted from heavy tail distributions like Lognormal and Pareto (continuous variables) and Zipf (discrete variables), with medium and large sizes. For medium size, we have observed also for Exponential and Gamma (continuous variables) and Geometric (discrete variables).

Finally, to check the emergence of the 4/3 power-law with observed data, in Fig.(\ref{fig:1}f), we compared sample values of skewness and kurtosis of daily precipitation$>0$, at Linkoeping-Malmllaett (Sweden), with a partition $n=50$ (black dots), and Eq.(\ref{equation7}) (blue line). With blue circles are indicated skewness and kurtosis couples having a value of the $R$ factor in the range [0.95, 1.05].
Blue line and blue circles are quite close. For this data, the condition given by Eq.(\ref{equation8}) is satisfied and the power-law relation emerges.

\section{Conclusions}
\begin{figure*}[b]
{\includegraphics[width=\textwidth]{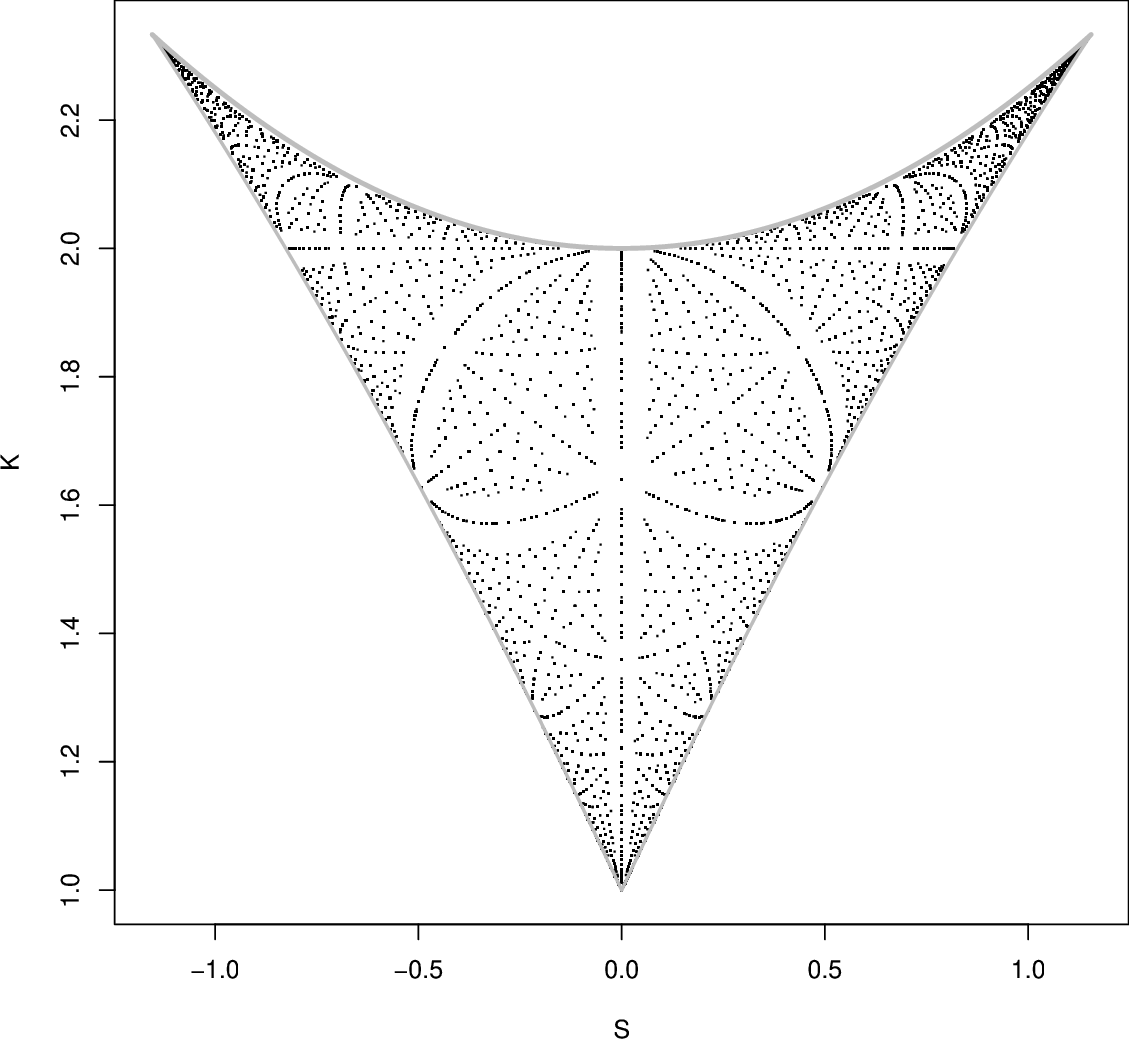}}
\caption{\label{fig:3} Skewness-kurtosis domain for $n=4$, obtained considering a number of synthetic samples $\Delta=10^8/n$ from the Poisson(rate=20) distribution.
In grey are reported also the upper limit given by Eq.(\ref{equation4}) and lower limit given by Table \ref{tab:table1}.}
\end{figure*}

\begin{figure*}[b]
    \centering

    \begin{subfigure}[b]{0.42\textwidth}
        \includegraphics[width=\textwidth]{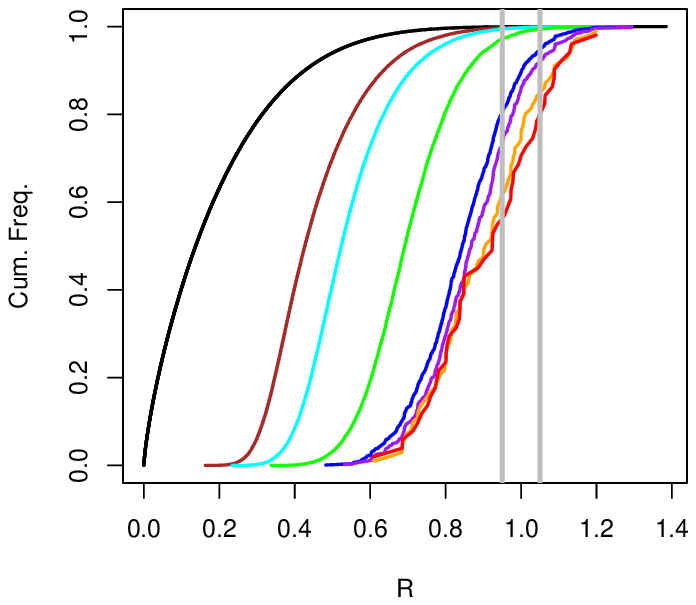}
        \caption{Gaussian, $n=100$}
    \end{subfigure}
    \hfill
    \begin{subfigure}[b]{0.42\textwidth}
        \includegraphics[width=\textwidth]{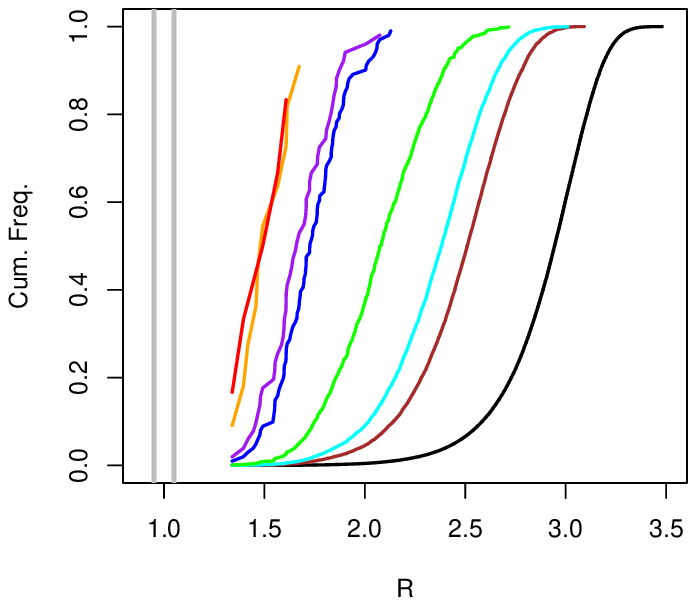}
        \caption{Exponential, $n=1000$}
    \end{subfigure}
    
    \vspace{2mm}

    \begin{subfigure}[b]{0.42\textwidth}
        \includegraphics[width=\textwidth]{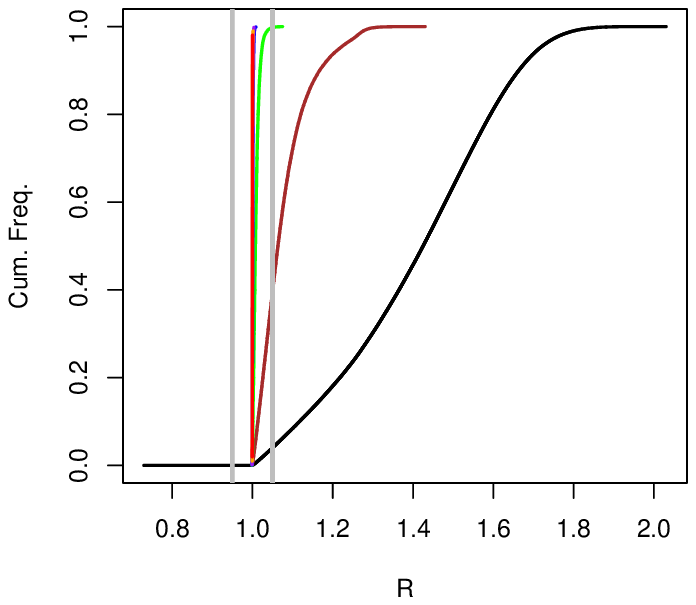}
        \caption{Lognormal, $n=100$}
    \end{subfigure}
    \hfill
    \begin{subfigure}[b]{0.42\textwidth}
        \includegraphics[width=\textwidth]{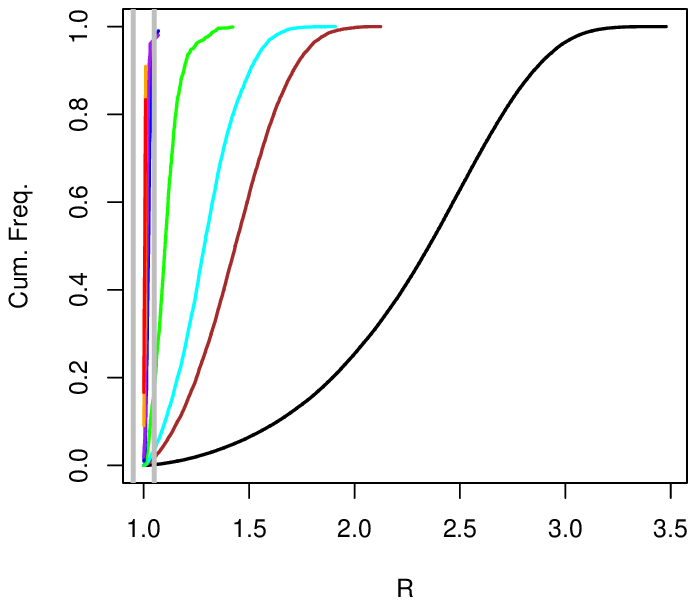}
        \caption{Pareto, $n=1000$}
    \end{subfigure}

    \caption{\label{fig:4} Cumulative frequency of $R$ (black line), and conditioned cumulative frequency of $R|S>s_q$ (where $s_q$ is the $q$-quantile of $S$), from $\Delta=10^8/n$ synthetic samples of size $n$ and distribution indicated in the caption of the panel. Brown for $q=0.9$, cyan for $q=0.95$, green for $q=0.99$, blue for $q=0.999$, purple for $q=0.9995$, orange for $q=0.9999$, and red for $q=0.99995$. The grey vertical lines delimit the range [0.95, 1.05] of $R$.}
\end{figure*}

Here, we have addressed two key and unexplored issues related to the behavior of sample skewness and kurtosis: 1) the one in correspondence of samples of small size ($n=4-9$), showing that the lower limit (like the upper limit) of $K$ is also a function of $n$, and providing the $n$-variable lower limit, and 2) the emergence of the 4/3 power-law behavior, which can be viewed as the extension of the Taylor Power-Law to a fourth-order theory. In particular, we assessed the conditions under which the 4/3 power-law behavior is observed.
Future analyses may aim to: (1) derive an analytical relation describing the behavior of sample kurtosis as a function of skewness and sample size; (2) investigate the symmetries and properties of the deltoid-shaped domain shown in Fig.(\ref{fig:3}); (3) assess similar issues in other metrics like L-skewness and L-kurtosis.    

%
\begin{table*}[htbp]
\caption{Left side of the lower limit (the right side is symmetrical respect y-axis) of the sample kurtosis varying $n$, $K=aS^2+bS+c$. In square parentheses the range of $S$ and round parentheses the values of $a$, $b$, and $c$, respectively.}
\label{tab:table1}
\centering
\begin{tabular}{|p{5.0cm} p{5.0cm} p{4.2cm}| c|}
\hline\noalign{\smallskip}
 & & \textbf{Left side} & $n$ \\
\noalign{\smallskip}\hline\noalign{\smallskip}
 & & $(-0.16930, -1.35019, 1)$ & 4 \\
 & & $[-1.1547, 0]$ & \\
\hline
 & $(-0.27282, -2.42886, 0.22056)$ & $(-0.5, 0, 1.25)$ & 5 \\
 & $[-1.5, -0.40825]$ & $[-0.40825, 0]$ & \\
\hline
 & $(-0.39975, -3.49778, -0.7778)$ & $(-0.72619, -1.2256, 1)$ & 6 \\
 & $[-1.78885, -0.71]$ & $[-0.71, 0]$ & \\
\hline
$(-0.46626, -4.39171, -1.85513)$ & $(-0.89555, -2.35544, 0.47801)$ & $(-1, 0, 1.16667)$ & 7 \\
$[-2.04124, -0.95394]$ & $[-0.95394, -0.28868]$ & $[-0.28868, 0]$ & \\
\hline
$(-0.49653, -5.12229, -2.91984)$ & $(-0.88454, -3.14719, -0.12097)$ & $(-1.30395, -1.18653, 1)$ & 8 \\
$[-2.26779, -1.155]$ & $[-1.155, -0.515]$ & $[-0.515, 0]$ & \\
\hline
$(-1.25168, -4.55704, -1.0906)$ & $(-1.58841, -2.43419, 0.57337)$ & $(-1.16882, 0, 1.12)$ & 9 \\
$[-1.325, -0.705]$ & $[-0.705, -0.234]$ & $[-0.234, 0]$ & \\
$(-0.47074, -5.59352, -3.83497)$ & & & \\
$[-2.47487, -1.325]$ & & & \\
\hline
\end{tabular}
\end{table*}

\begin{table*}[b]
\caption{\label{tab:table2}%
Emergence (Y) or not (N) of the power-law behavior for different values of the sample size $n$ and different continuous (Gaussian, Exponential, Lognormal, Gamma and Pareto) and discrete (Binomial, Negative Binomial, Poisson, Geometric and Zipf) distributions. \textsuperscript{'}*\textsuperscript{'} indicates cases illustrated in Fig.4.}
\begin{tabular}{|p{5.5cm}|p{1.7cm}|p{1.7cm}|p{1.7cm}|}
\hline
\textrm{Probability Distribution}&
\textrm{$n=4$}&\textrm{$n=100$} & $n=1000$\\
\hline
 $Gaussian(mean=0,sd=1)$ & N & N* & N\\
 \hline
  $Exponential(rate=1)$ & N & Y & N*\\
  \hline
  $Lognormal(meanlog = 0, sdlog = 1)$  & N & Y* & Y\\
    \hline
 $Gamma(shape=2, rate=1)$ & N &  Y & N\\
  \hline
  $Pareto(shape=5, scale=1)$ & N & Y & Y* \\
   \hline\hline
 $Binomial(trials=20, p=0.8)$ & N &  N & N\\
  \hline
 $Negative Binomial(trials=20, p=0.8)$ & N & N & N\\
  \hline
 $Poisson(rate=20)$ & N & N & N\\
  \hline
$Geometric(p=0.8)$ & N & Y & N\\
  \hline
  $Zipf(shape=5, xmin=1)$  & N & Y & Y\\
  \hline
\end{tabular}
\end{table*}

\section*{Data Availability Statement}
The air humidity data for Toulouse (France) are available at \url{https://data.toulouse-metropole.fr/explore/dataset/}.  
The precipitation data for Linköping-Malmslätt (Sweden) can be accessed at \url{https://www.ecad.eu/dailydata/index.php}.  
The Rakaia and Brahmaputra datasets (i.e., the number of wet channels for each cross section) supporting the findings of this study are available from S.D.B. upon reasonable request.

\section*{Acknowledgements}
This study was carried out within the RETURN project (Multi-risk science for resilient communities under a changing climate, \url{https://www.fondazionereturn.it/}) with Extended Partnership, which received funding from the European Union Next-GenerationEU, National Recovery and Resilience Plan—NRRP, Mission 4, Component 2, Investment 1.3—D.D. 1243 2/8/2022, PE0000005.

\bibliographystyle{spphys}       
\bibliography{epj}   

\end{document}